# Spin Wave Electromagnetic Nano-Antenna Enabled by Tripartite Phonon-Magnon-Photon Coupling

Raisa Fabiha[1], Jonathan Lundquist[1], Sudip Majumder[2], Erdem Topsakal[1], Anjan Barman[2] and Supriyo Bandyopadhyay[1]

**ABSTRACT**

**We investigate tripartite coupling between phonons, magnons and photons in a periodic array of elliptical magnetostrictive nanomagnets delineated on a piezoelectric substrate to form a two-dimensional two-phase multiferroic crystal. A surface acoustic wave (phonons) of 5 – 35 GHz frequency launched into the substrate causes the magnetizations of the nanomagnets to precess at the frequency of the wave, giving rise to spin waves (magnons). The spin waves, in turn, radiate electromagnetic waves (photons) into the surrounding space at the surface acoustic wave frequency. Here, the phonons couple into magnons, which then couple into photons. This tripartite phonon-magnon-photon coupling is exploited to implement an extreme sub-wavelength electromagnetic antenna whose measured radiation efficiency and antenna gain exceed the theoretical limits for traditional antennas by more than two orders of magnitude at some frequencies. Micro-magnetic simulations are in excellent agreement with experimental observations and provide insight into the spin wave modes that couple into radiating electromagnetic modes to implement the antenna.**

[1] Department of Electrical and Computer Engineering, Virginia Commonwealth University, Richmond, VA 23284, USA

[2] Department of Condensed Matter Physics and Material Sciences, S. N. Bose National Center for Basic Sciences, Block JD, Sector III, Salt Lake, Kolkata 700 106, India

# I. INTRODUCTION

Electromagnetic antennas are actuated by passing an alternating current (or an electromagnetic wave) through a metallic or non-metallic element, which causes the latter to radiate electromagnetic waves of the same frequency as the excitation into the surrounding medium. A serious drawback of this implementation is that if the physical dimensions of the antenna are much smaller than the wavelength of the radiation they emit, then the radiation efficiency and the gain become very poor. If the antenna impedance does not exceed that of free space (377 ohms) [which is usually the case], then the radiation efficiency is limited to $\sim A/\lambda^2 \left( A << \lambda^2 \right)$ [1], where $A$ is the emitting area of the antenna and $\lambda$ is the emitted wavelength, while the gain is limited to $A/(2\pi\lambda)^2 + \left( \sqrt{A}/\pi\lambda \right)$ [2]. This makes it extremely difficult to miniaturize antennas (making them much smaller than the wavelength) without sacrificing efficiency and gain.

Recently, we overcame the above limit on the radiation efficiency with a magneto-elastic antenna [3]. The antenna consisted of elliptical magnetostrictive nanomagnets elastically coupled to an underlying piezoelectric substrate. A low frequency acoustic wave (144 MHz) launched in the substrate periodically strained the nanomagnets, causing their magnetizations to rotate through small angles because of the inverse magnetostriction (Villari) effect. The time-varying magnetizations then emitted electromagnetic waves at the excitation frequency of 144 MHz ($\lambda = 2$ m) with an efficiency that exceeded the $A/\lambda^2$ limit by more than five orders of magnitude, but *only at low frequencies* (well below 1 GHz) since the magnetization oscillations due the Villari effect could not occur with sub-ns period.

In this work, we demonstrate an electromagnetic antenna operating on a different principle. It radiates efficiently at much higher frequencies (1-30 GHz). A surface acoustic wave (SAW) of ~GHz frequency excites spin waves (of the same frequency) in the nanomagnets via the coupling of the phonons in the SAW to magnons in the spin wave [4]. The generated spin waves can then couple to electromagnetic radiative modes via magnon-photon coupling, thereby radiating electromagnetic waves into the surrounding medium. This implements an electromagnetic antenna. We observe high radiation efficiency and antenna gain (more than two orders of magnitude larger than the theoretical limits for traditional antennas) at two frequencies (5 and 14 GHz) where two conditions are satisfied. First, we are able to couple energy efficiently from a microwave source into the SAW at these frequencies owing to good impedance matching, and second, these SAW are able to excite high amplitude (high power) spin waves in the nanomagnets because of relatively efficient coupling between the surface acoustic and spin waves (phonon-magnon coupling) at these frequencies. Micromagnetic simulations confirm that indeed high amplitude spin waves are excited by SAW of these frequencies in the nanomagnets. The same simulations provide the power and phase profiles of the spin waves revealing intriguing variations in space.

We also observed that the gain, efficiency, spin wave profiles, etc. depend on the direction of propagation of the SAW with respect to the nanomagnet geometry, i.e. whether the propagation is along the major or minor axes of the nanomagnets. This is not surprising since the major axis is the magnetic easy axis and the minor axis is the magnetic hard axis. Additionally, the edge-to-edge separation between the nanomagnets (and hence the degree of dipole coupling between neighbors) is also different along the two directions, which spawns the anisotropy.

## II. RESULTS AND DISCUSSION

**Spin wave nano-antennas**

Interaction between spin waves and SAW have been studied in a wide variety of scenarios such as parametric amplification of spin waves via coupling to SAW [4, 5]. Coupling between magnons and phonons was also studied, aided by extensive simulations [6-8]. Strong coupling leads to the formation of a hybridized magnon-phonon quasi-particle called a magnon polaron [9, 10]. Other studies of coupling between spin waves and SAW have also been reported [11-17]. Here, we have harnessed this interaction to implement a novel antenna whose gain and radiation efficiency exceed the theoretical limits by two orders of magnitude or more.

The spin-wave nano-antennas reported here are fabricated by depositing elliptical cobalt nanomagnets of major axis dimension ~120 nm and minor axis dimension ~ 110 nm on a $LiNbO_3$ substrate, using electron beam lithography to pattern a resist, followed by resist development, electron beam evaporation of metal on to the patterned surface, and lift-off. The edge-to-edge separation between the nanomagnets along the direction of the major axis is ~25 nm and along the direction of the minor axis is ~60 nm, as shown in the scanning electron micrograph (SEM) in Fig. 1(a). The nanomagnets are arranged into 200 squares, each containing 80× 80 nanomagnets. The SEM of a square can be found in the supplementary material. The area covered by the nanomagnets, which is the antenna area $A$, is 10,505 μm$^2$.

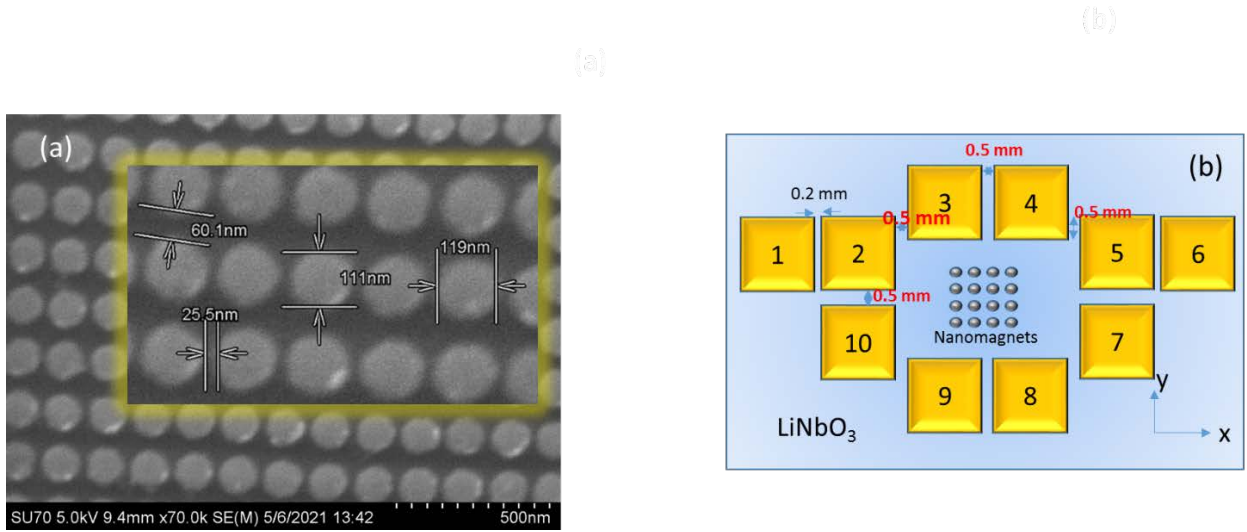

Fig. 1: (a) Scanning electron micrograph of elliptical Co nanomagnets deposited on a LiNbO3 substrate. The inset shows the dimensions. (b) The antenna configuration with electrodes for launching surface acoustic waves that subject the nanomagnets to time-varying strain at the frequency of the launched wave. Fig. 1(b) is not to scale.

Aluminum electrodes are deposited around the nanomagnet assembly (using optical lithography) for launching SAW in the substrate in different directions, as shown in Fig. 1(b). These electrodes are not interdigitated transducers (IDTs) since IDTs are frequency filters that are efficient only at the resonant frequency and inefficient at all other frequencies. Since we wish to study the antenna radiation over a broad spectrum of frequencies, and not just at one frequency, we needed to forego IDTs and use solid electrodes, which are not narrow band filters and hence work over a broad range of frequencies. The surface acoustic waves (SAWs) launched by these electrodes are of course not the traditional Rayleigh, Sezawa, Lamb or Love modes, but they nonetheless generate time varying strain on the nanomagnets, which, in turn, excites spin waves in the nanomagnets. We call them "surface" acoustic waves since the acoustic wavelength (~0.5 μm) is much smaller than the substrate thickness (0.5 mm). Since the acoustic wave will decay into the substrate with a characteristic length that is about the wavelength, it will be mostly constrained to the surface. Optically generated SAW (generated by ultrashort laser pulses) in a piezoelectric substrate are also known to produce spin waves in magnetostrictive nanomagnets, and that has been studied extensively in the past using, primarily, time-resolved magneto-optical Kerr effect microscopy [6-9]. Here, we do not use optical generation, but use the more traditional modality of generating them with an electrical (microwave) source.

In our samples, we apply a microwave frequency voltage between electrodes 3 and 4, as well as between electrodes 5 and 7. In the former case, the SAW would propagate primarily along the minor axis of the elliptical nanomagnets, and in the latter case, along the major axis.

SAWs of frequencies in the range of 1-30 GHz are launched in the LiNbO3 substrate by applying 1 dbm (1 mW) of power from the microwave source between a pair of electrodes. As mentioned earlier, we choose two different pairs (3, 4) and (5, 7) in Fig. 1(b), which would launch SAWs propagating in two different directions. We label the two cases as "orientation 1" and "orientation 2", respectively. We test multiple (nominally identical) samples and a control sample for background subtraction. The control sample is identical to a real sample, except it has no nanomagnet.

**Microwave testing of spin wave nano-antennas**

Using a network analyzer, we first measure the scattering parameter $S_{11}$ as a function of frequency (1-40 GHz). This parameter measures the reflection back into the source due to impedance mismatch between the device and the source. In Fig. 2(a) and 2(b), we show the $S_{11}$ versus frequency plots for a representative sample in the two orientations 1 and 2. There are sharp dips at some frequencies (5, 14, 30 and 35 GHz for this sample) where the reflection is small (owing to good impedance matching) and at these frequencies, a large fraction of the incident power will be coupled into the SAWs that are launched into the substrate. We therefore measure the radiated electromagnetic power when the source frequency is one of these four frequencies. For the sake of comparison, we also measure the radiated power at a few other frequencies where the reflection ($S_{11}$) is high and coupling of energy from the source into the SAW is relatively poor. The radiated power is expectedly lower at these frequencies.

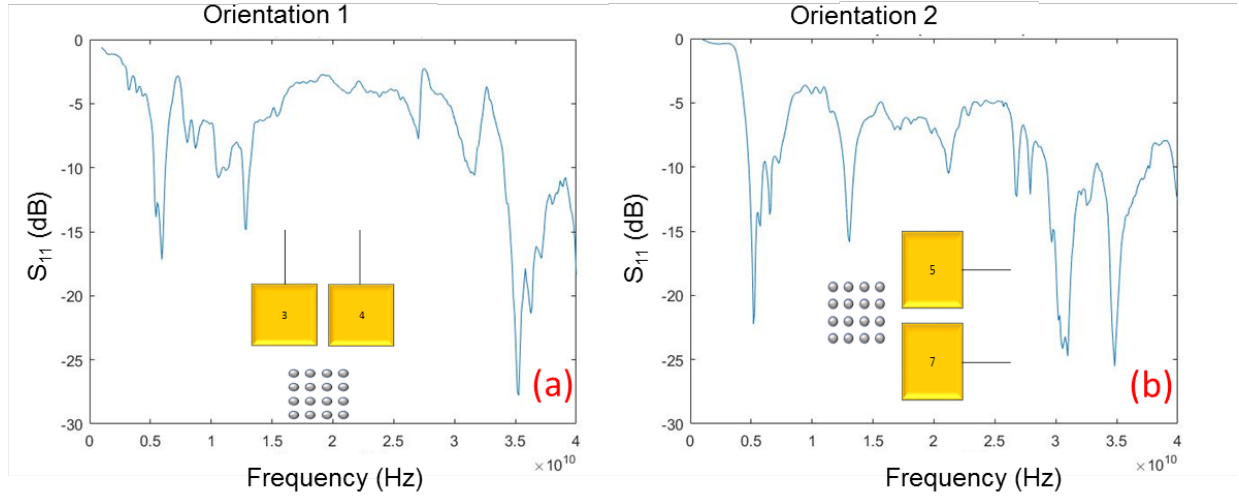

Fig. 2: The scattering parameter $S_{11}$ for a representative sample when excited in two different orientations (corresponding to two different directions of propagation of the SAW), measured in the frequency range 1 – 40 GHz. (a) Orientation 1 where the direction of propagation of the surface acoustic wave is primarily parallel to the minor axes (hard axes) of the nanomagnets. (b) Orientation 2 where the direction of propagation of the surface acoustic wave is primarily parallel to the major axes (easy axes) of the nanomagnets.

A calibrated horn antenna, placed at a fixed distance from the sample, measures the radiated electromagnetic power, as shown in Fig. 3(a). Radio frequency absorbers surround the entire setup during measurement. The measured noise floor was -90 dB. This distance is greater than four times the wavelength of the emitted electromagnetic wave to ensure that we are measuring the far field radiation. The horn antenna feeds into a digital spectrum analyzer, which provides a measurement of the received power as a function of frequency. Fig. 3(b) shows a representative spectrum of the received power from one sample when the excitation frequency is 5 GHz. We always measure the received power from both a real sample and the control sample; the actual received power from a sample is the difference of the two. We find that the actual received power always peaks at the frequency of microwave excitation, i. e. the frequency of the SAW launched into the piezoelectric substrate. However, we also see peaks (sidebands) at other frequencies as well, as can be seen distinctly in Fig. 3. Some, *but not all*, of these sidebands can be spurious (e.g. due to extraneous radiation sources).

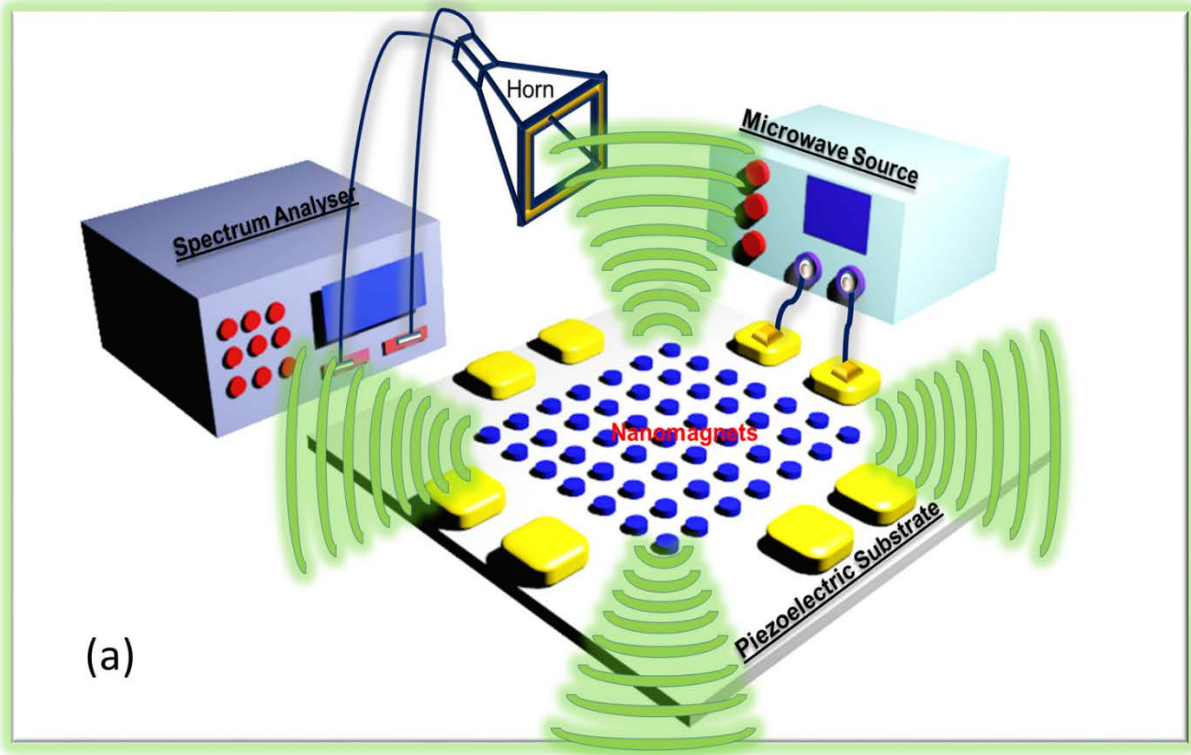

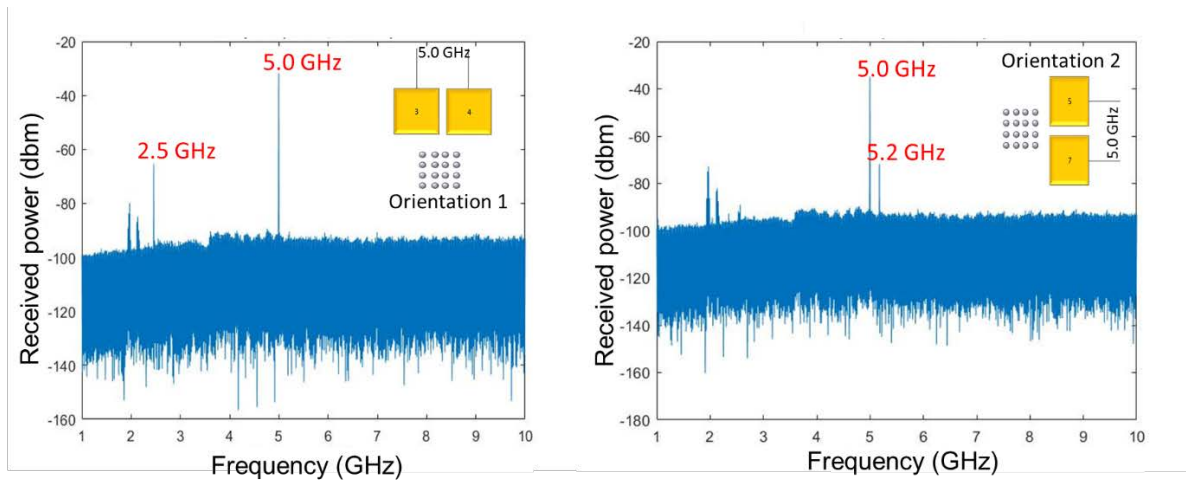

Fig. 3: (a) Antenna measurement setup. (b) Spectrum of received power at the horn antenna when the antenna was placed 36.8 cm from a sample. The excitation frequency is 5 GHz, and that is where we obtain the dominant peak in the received power spectrum. The spectra are shown for both orientations used in launching the SAW. The two broad peaks at 2.0 and 2.2 GHz are probably due to spurious sources that could not be eliminated. Because they are spurious, they show up in both orientations. The peak at 2.5 GHz is a sideband associated with a non-resonant spin wave (spin wave excited at a frequency different from the SAW frequency of 5 GHz) as shown by later simulations. The 5.2 GHz peak could not be explained with simulations and is possibly also from spurious sources.

Neglecting polarization effects, we calculate the antenna gain at any given frequency from the relation $P_r = G_t G_r P_t \left(\frac{\lambda}{4\pi d}\right)^2$ [18] where $P_r$ is the actual received power at the horn antenna (power received from the real sample minus power received from the control sample), $P_t$ is the power coupled into the sample from the microwave source, $G_r$ is the horn antenna gain (measured with a second horn antenna placed at the same distance from the first as the sample), $G_t$ is the (directionally averaged) spin-wave antenna gain, $d$ is the distance between the sample and the horn antenna, and $\lambda$ is the wavelength. At any frequency, the quantity $P_t = (1-r)$ mW, where $r$ is the reflection coefficient calculated from the parameter $S_{11}$ (see supplementary material) at that frequency. We define the 'radiation efficiency' as $\eta = P_r/P_t$.

In Table 1, we list the measured antenna gain $G_t$ for a representative sample excited in the two different orientations (two different directions of propagation of the SAW) at various frequencies, some of which (5, 14, 30 and 35 GHz) correspond to dips in the $S_{11}$ spectra, while the others do not. We also list the value of the theoretical maximum gain $G_{max}^{theor}$ at any frequency for a traditional antenna calculated using the relation given in ref. [2]. In Table 2, we list the radiation efficiency $\eta$ for the same sample for the two different orientations at various frequencies. We also list the theoretical efficiency limit $A/\lambda^2$ at each frequency.

**Table 1: Measured spin-wave nano-antenna gain $G_t$ for a representative sample excited in two different orientations (two different directions of propagation of the SAW)**

| Frequency (GHz) | Sample 3 Orientation 1 | Sample 3 Orientation 2 | $G_{max}^{theor}$ |
|---|---|---|---|
| 5 | 0.356 | 0.116 | $5.5 \times 10^{-4}$ |
| 10 | 0.0874 | | 0.0011 |
| 13 | | 0.0725 | 0.0014 |
| 14 | | 0.544 | 0.0015 |
| 15 | 0.149 | 0.121 | 0.0016 |
| 18 | 0.00138 | | 0.0019 |
| 30 | | 0.047 | 0.0032 |
| 35 | | 0.0135 | 0.0038 |

**Table 2: Measured spin-wave nano-antenna radiation efficiency η for a representative sample excited in two different orientations (two different directions of propagation of the SAW)**

| Frequency (GHz) | Sample 3 Orientation 1 | Sample 3 Orientation 2 | $A/\lambda^2$ |
|---|---|---|---|
| 5  | $3.78 \times 10^{-4}$ | $1.22 \times 10^{-4}$ | $2.9 \times 10^{-6}$ |
| 10 | $2.92 \times 10^{-5}$ |                       | $1.16 \times 10^{-5}$ |
| 13 |                       | $1.41 \times 10^{-5}$ | $1.97 \times 10^{-5}$ |
| 14 |                       | $1.42 \times 10^{-4}$ | $2.28 \times 10^{-5}$ |
| 15 | $4.46 \times 10^{-5}$ | $3.42 \times 10^{-5}$ | $2.61 \times 10^{-5}$ |
| 18 | $1.32 \times 10^{-7}$ |                       | $3.76 \times 10^{-5}$ |
| 30 |                       | $7.36 \times 10^{-7}$ | $1.04 \times 10^{-4}$ |
| 35 |                       | $9.89 \times 10^{-8}$ | $1.42 \times 10^{-4}$ |

Looking at Table 1, we see that the frequencies where we obtain high antenna gain are at 5 and 14 GHz. At those frequencies, the gains are 647 and 363 times higher than the theoretical limits, respectively. Similarly, looking at Table 2, we see that the radiation efficiency is high at the same two frequencies and in the same orientations. The ratio of the measured radiation efficiency $\eta$ to the quantity $A/\lambda^2$ is 130 and 6 at these two frequencies (5 and 14 GHz). All these ratios are calculated when the detecting horn antenna is at a distance of ~36 cm from the sample (which is 6 times the wavelength of free space electromagnetic radiation at 5 GHz and nearly 17 times the wavelength at 14 GHz).

We note two other important features: a clear orientation dependence (anisotropy) whereby the direction of propagation of the SAW with respect to the nanomagnets' easy axes makes a difference, and that the gain and efficiency are not large at every frequency where $S_{11}$ has a dip (i.e. power is efficiently coupled into the SAW from the microwave source). There are pronounced dips at 35 GHz and 30 GHz (the latter in orientation 2), but both the gain and the radiation efficiency are poor at those two higher frequencies, especially the efficiency. This tells us that mere high transfer of power into the SAW (or generation of large amplitude SAW) *does not guarantee* high amount of electromagnetic radiation. This is because coupling of power from the SAW to the spin wave (phonon-magnon coupling) and coupling of power from the spin wave into the electromagnetic wave (magnon-photon coupling) also matter. The coupling efficiency depends on frequency. Micromagnetic simulations show that the amplitudes of spin waves excited in the nanomagnets are strongly frequency-dependent and that explains the observed frequency dependence of the gain and radiation efficiency. For example, we find that high power spin waves are excited in the nanomagnets at 5 GHz, but the power is much lower at 35 GHz, which explain why we observe strong radiation at 5 GHz, but not at 35 GHz.

**Micro-magnetic simulations**

Micromagnetic simulations are carried out with Object Oriented Micromagnetic Framework (OOMMF) software, which solves the Landau-Lifshitz-Gilbert equation with effective magnetic fields due to shape anisotropy, dipole coupling between neighbors, exchange, and time-varying strain. The simulation is carried out on a $7 \times 7$ array of nanomagnets, discretizing the samples into rectangular prisms of dimensions $2 \times 2 \times 6$ nm$^3$. The time varying strain in the nanomagnets acts like a time-varying magnetic field given by $H_{SAW}(t) = \frac{3\lambda_s \sigma(t)}{2\mu_0 M_s}$, where $\mu_0$ is the magnetic permeability of free space, $M_s$ is the saturation magnetization of the nanomagnets (~1 MA/m), $\lambda_s$ is the saturation magnetostriction of Co, and $\sigma(t) = \sigma_0 \sin(2\pi f t)$ is the sinusoidal time varying stress due to the SAW [19]. The stress frequency $f$ is the same as that of the SAW. The amplitude $\sigma_0$ is calculated from the relation [20] $\sigma_0 = \sqrt{2PZ_0}$; $Z_0 = \sqrt{C_{11}\rho}$ where $P$ is the power in the SAW per unit area (normal to the direction of propagation), $Z_0$ is the characteristic acoustic impedance, $C_{11}$ is the first diagonal element of the elasticity tensor, and $\rho$ is the mass density of the LiNbO$_3$ substrate. The cross-section area normal to the direction of propagation of the SAW is the penetration depth of the SAW times the width of the electrodes. The penetration depth is approximately the wavelength of the SAW, which varies with frequency, but since we can only carry out an order estimate, we will neglect the frequency dependence and calculate the wavelength at 10 GHz (which is in the midrange of the frequencies we have measured at). The velocity of acoustic wave in solids is typically between 3 km/s – 6 km/s and hence we will assume an average velocity of 4.5 km/s, which will yield an average penetration depth of 0.45 μm. Since the nanomagnet thickness (~6 nm) is much smaller than the penetration depth of the SAW, we will assume that all of the strain due to the SAW generated in the substrate is transferred to the nanomagnets.

The width of the electrode is 2 mm and therefore the cross-sectional area is $9 \times 10^{-10}$ m$^2$. The power coupled into the SAW is $P_t = (1-r)$ mW, where $r$ is the reflection coefficient calculated from the parameter S$_{11}$. Again, we will ignore the frequency dependence of $r$ and take its average value at the critical frequencies of 5, 14, 30 and 35 GHz, i.e. $r = $ ~0.01 (for S$_{11}$ = -20 dB). Hence $P_t = 0.99$ mW and $P = 1.1 \times 10^6$ W/m$^2$. For LiNbO$_3$, $C_{11} = 202$ GPa and $\rho = 4650$ kg m$^{-3}$. This yields $Z_0 = 9.7 \times 10^8$ N-S-m$^{-3}$. This yields $\sigma_0 = 46$ MPa. The reported saturation magnetostriction of Co has a wide range between 30 ppm and 150 ppm [21] and we will assume it to be 70 ppm. Hence the amplitude of $H_{SAW}(t)$ is 3788 A/m = 46 Oe.

We assume that the major axis of the nanomagnets is along the $x$-direction and the minor axis is along the $y$-direction, as shown in Fig. 1(b). The micromagnetic simulations yield the magnetization components $m_x(x,y,z,t), m_y(x,y,z,t), m_z(x,y,z,t)$ as a function of space and time. From these quantities, we obtain the spatially averaged magnetization components $M_x(t), M_y(t), M_z(t)$ as a function of time. We then compute the fast Fourier transforms (FFT) of the oscillations to obtain the spin wave spectra. The power and phase profiles of the dominant spin wave peak in the spectra are calculated with an in-house simulator DOTMAG described in ref. [22].

In Fig. 4, we show the calculated oscillations in $M_y$ (magnetization component along the minor axes of the nanomagnets) when a 5 GHz SAW is generated by applying a voltage between contacts 5 and 7 (orientation 2). The Fourier transforms of these oscillations are shown in both linear and log-linear scales. We observe that the dominant peak occurs at the driving frequency of 5 GHz, but there are also much weaker sidebands at 2.5 GHz, 12.5 GHz and 15 GHz. The power and phase profiles of the spin waves associated with the dominant peak show that the wave is quantized along the major axis of the ellipse (which is also the direction of SAW propagation) with quantization number = 2.

In Fig. 5, we show the calculated oscillations in $M_x$ (magnetization component along the major axis of the ellipse) and $M_z$ (normal-to-plane component) and their Fourier transforms when the SAW frequency is 5 GHz and the sample is excited in orientation 2. Interestingly, the dominant frequency in the $M_x$ oscillation spectrum is *twice* the driving frequency (10 GHz), although there is a weaker sideband at the driving frequency of 5 GHz. The dominant frequency in the $M_z$ oscillation spectrum, on the other hand, is the driving frequency of 5 GHz, and there are weaker sidebands at 2.5, 12.5 and 15 GHz. We note that the power in the dominant frequency component of $M_y$ oscillations is *almost one (two) order of magnitude larger* than that in the case of $M_x$ ($M_z$) oscillations. Hence, the spin wave is composed mostly of oscillations in the component of the magnetization *along the minor axis* when the SAW is propagating along the major axes of the elliptical nanomagnets. The dominant peak is unambiguously at 5 GHz, which matches the spectrum in Fig. 3(b). We see no hint of the peak at 5.2 GHz in the simulations and therefore we believe that it is spurious.

In Fig. 6, we show the calculated oscillations in $M_y$ (magnetization component along the minor axes of the nanomagnets) when a 5 GHz SAW is generated by applying a voltage between contacts 3 and 4 (orientation 1) and its Fourier transforms in both linear and log-linear scales. We observe that the dominant peak occurs at *one-half* of the driving frequency of 5 GHz, i. e. at 2.5 GHz, but there are also much weaker sidebands at 1.0 GHz, 5.0 GHz, 7.5 GHz and 12.5 GHz.

This explains the 2.5 GHz peak that we see in orientation 1 in Fig. 3(b), but not in orientation 2. This oscillation is *not* at the driving SAW frequency and hence must be *intrinsic* to the array. It is spawned by the internal magnon dynamics determined by array parameters, namely the demagnetization field and the dipole coupling field, which, in turn are determined by the magnet dimensions, saturation magnetization and the edge-to-edge separation between neighboring nanomagnets. Since the demagnetization and dipolar field components along the direction of SAW propagation is different in the two 'orientations', we see the effect of the intrinsic mode on radiated power in orientation 1, but not in orientation 2.

In Fig. 7, we show the calculated oscillations in $M_x$ (magnetization component along the major axis of the ellipse) and $M_z$ (out-of-plane component) and their Fourier transforms when the SAW frequency is 5 GHz and the sample is excited in orientation 1. The dominant peak in the spectrum of $M_x$ oscillation occurs at the driving frequency of 5 GHz, with much weaker side bands at 1.0 GHz, 2.5 GHz, 10 GHz and 15 GHz. The dominant peak in the spectrum of $M_z$ oscillation, however, does not occur at the driving frequency of 5 GHz, but occurs at 2.5 GHz, which is similar to the case of the $M_y$ component. Again, this is the signature of the intrinsic spin wave mode stimulated in orientation 1, but not orientation 2. There are weaker peaks at 1.0, 5.0, 7.5 and 12.5 GHz. Once again, the power in the dominant frequency component of $M_y$ oscillations is *almost one (two) order of magnitude larger* than that in the case of $M_x$ ($M_z$) oscillations. Hence, as in the case of orientation 2, the spin wave excited in orientation 1 is composed mostly of oscillations in the component of magnetization *along the minor axes* of the elliptical nanomagnets. Therefore, regardless of the direction of SAW propagation, the power in the spin waves is mostly concentrated in the oscillation of the magnetization component along the minor axes of the nanomagnets, which is the magnetic hard axis.

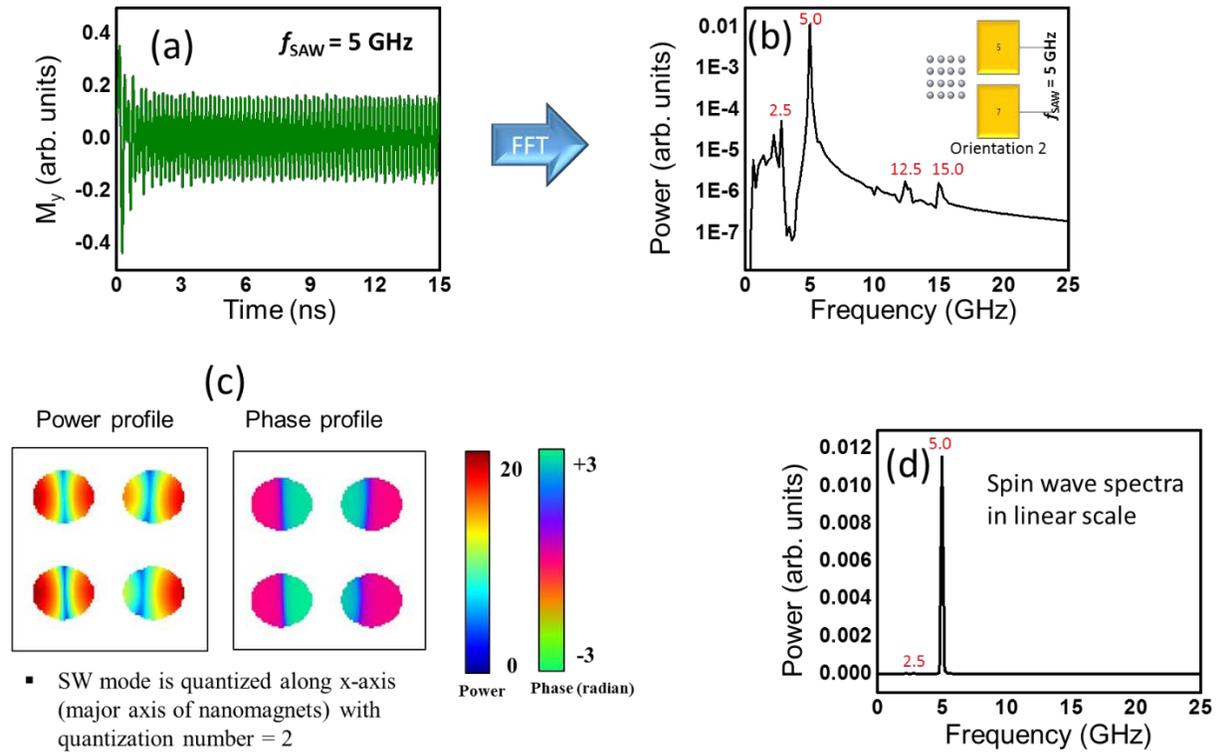

Fig. 4: (a) Calculated oscillations in the magnetization component parallel to the minor axis of the elliptical nanomagnets (hard axis) when the exciting SAW frequency is 5 GHz. The sample is excited in orientation 2 where the direction of propagation of the SAW is along the major axes of the elliptical nanomagnets. (b) The power spectrum of the oscillations in log linear scale where the dominant peak is at the excitation frequency. The next dominant peak has a power amplitude more than 2 orders of magnitude smaller. (c) The power and phase profiles of the spin waves in the 5 GHz (dominant frequency) mode showing that the wave is quantized along the major axis (also direction of propagation of the SAW) with quantization number = 2. (d) The power spectrum of the oscillations in linear scale.

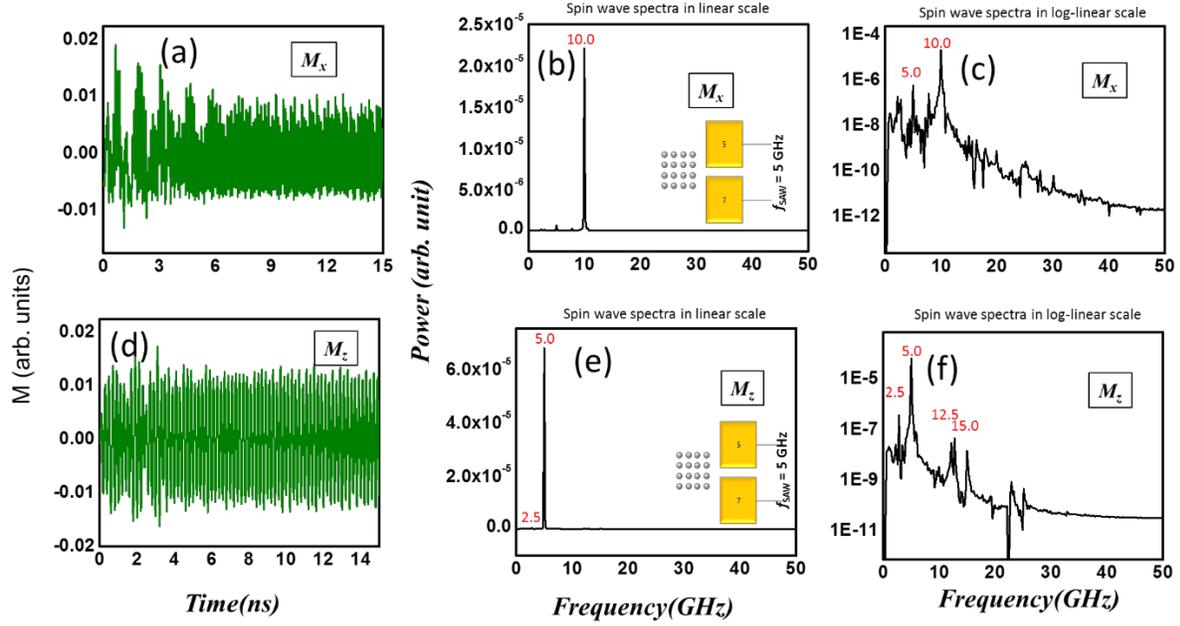

Fig. 5: (a) Calculated oscillations in the magnetization component parallel to the major axis of the elliptical nanomagnets (easy axis) when the exciting SAW frequency is 5 GHz. The sample is excited in orientation 2 where the direction of propagation of the SAW is along the major axes of the elliptical nanomagnets. (b) The power spectrum of the oscillations in linear scale where the dominant peak is at *twice* the excitation frequency. (c) The power spectrum in log linear scale showing some side bands. The next dominant peak is at the excitation frequency of 5 GHz and has a power amplitude more than 2 orders of magnitude smaller than the dominant peak. The power in the dominant oscillation mode of $M_x$ (at 10 GHz) is more than 2 orders of magnitude smaller than that in the dominant oscillation mode of $M_y$ (at 5 GHz) shown in Fig. 4. (d) Oscillations in the out-of-plane magnetization component. (e) The power spectrum of the oscillations in linear scale where the dominant peak is at the excitation frequency of 5 GHz. (f) The power spectrum in log linear scale showing some side bands. Again, the power in the dominant oscillation mode of of $M_z$ (at 5 GHz) is more than 2 orders of magnitude smaller than the power in the dominant oscillation mode of $M_y$ at 5 GHz shown in Fig. 4.

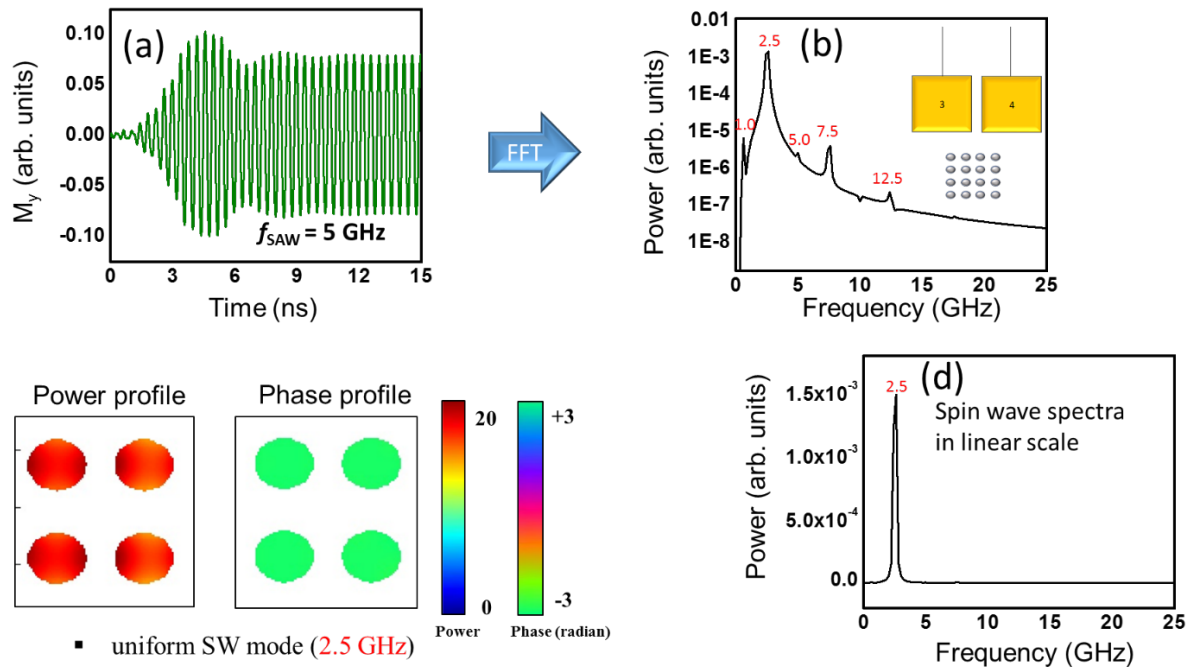

Fig. 6: (a) Calculated oscillations in the magnetization component parallel to the minor axis of the elliptical nanomagnets (hard axis) when the exciting SAW frequency is 5 GHz. The sample is excited in orientation 1 where the direction of propagation of the SAW is along the minor axes of the elliptical nanomagnets. (b) The power spectrum of the oscillations in log linear scale where the dominant peak is at one-half the excitation frequency (2.5 GHz). The next dominant peak has a power amplitude more than 2 orders of magnitude smaller. (c) The power and phase profiles of the spin waves in the 2.5 GHz (dominant frequency) mode showing that the profiles are uniform. (d) The power spectrum of the oscillations in linear scale.

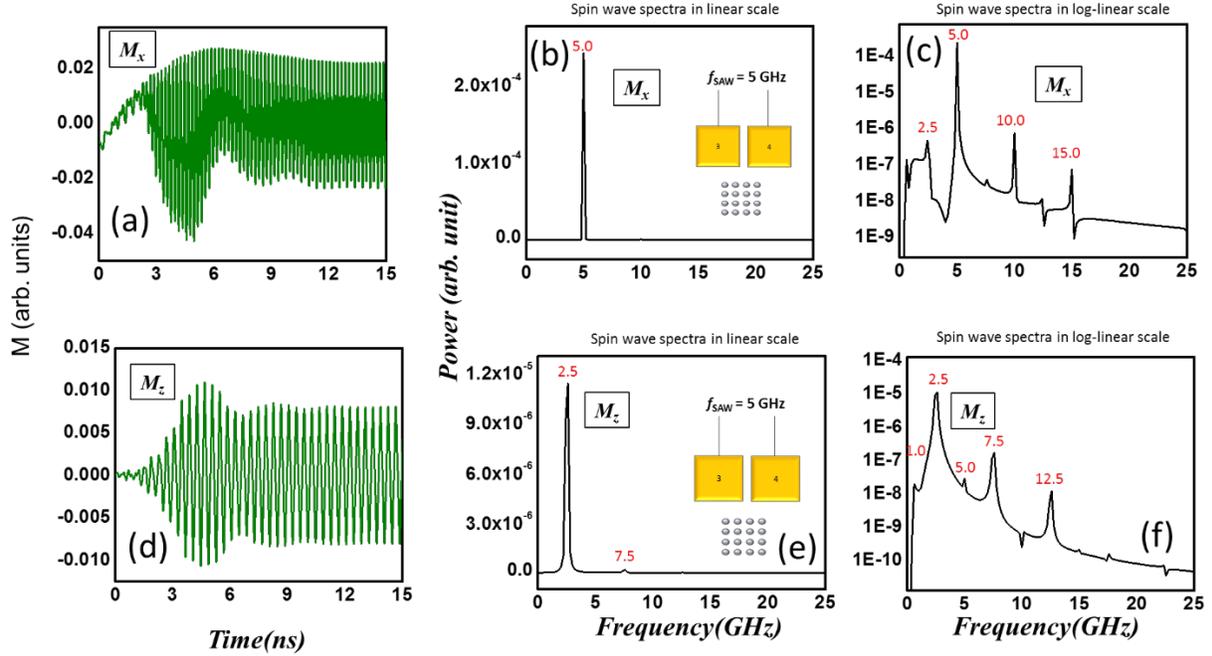

Fig. 7: (a) Calculated oscillations in the magnetization component parallel to the major axis of the elliptical nanomagnets (easy axis) when the exciting SAW frequency is 5 GHz. The sample is excited in orientation 1 where the direction of propagation of the SAW is along the minor axes of the elliptical nanomagnets. (b) The power spectrum of the oscillations in linear scale where the dominant peak is at the excitation frequency. (c) The power spectrum in log linear scale showing some side bands. The next dominant peaks are at one-half and twice the excitation frequency of 5 GHz and have power amplitudes more than 2 orders of magnitude smaller than that of the dominant peak. The power in the dominant oscillation mode of $M_x$ at 5 GHz is more than 2 orders of magnitude smaller than that in the dominant oscillation mode of $M_y$ (at 2.5 GHz) shown in Fig. 6. (d) Oscillations in the out-of-plane magnetization component. (e) The power spectrum of the oscillations in linear scale where the dominant peak is at one-half of the excitation frequency. (f) The power spectrum in log linear scale showing some side bands. Again, the powers in the dominant and side band peaks of $M_z$ oscillations are more than 2 orders of magnitude smaller than the power in the dominant oscillation mode of $M_y$ at 2.5 GHz shown in Fig. 6.

These results reveal many interesting features. First, the spin wave power is always mostly in the oscillation of the magnetization component along the minor axis (hard axis) of the nanomagnets, regardless of whether that axis is parallel or perpendicular to the direction of SAW propagation. Second, there is a strong orientation dependence since in orientation 1, the dominant peak (peak with most power) occurs at one half of the excitation frequency (at 2.5 GHz) due to an intrinsic mode, whereas in orientation 2, it occurs at the excitation frequency (5 GHz) due to the extrinsic mode stimulated by the SAW. Both intrinsic and extrinsic modes were observed in ref. [4]. Third, the dominant peak power is about one order of magnitude higher in orientation 2 (SAW propagating along the major axis of the nanomagnets) than in orientation 1. The radiated power spectra and simulated results at 14, 30 and 35 GHz (where $S_{11}$ has dips and therefore surface acoustic waves of these frequencies can be launched in the substrate efficiently) are shown in the Supplementary Material.

Whenever the dominant peak is at a frequency *different* from the excitation frequency of 5 GHz, the dominant mode is an intrinsic mode. The array has its own intrinsic spin wave modes determined by the nanomagnet and array parameters [4], as stated previously. When a SAW excites an intrinsic mode, hybridization of the intrinsic mode with the extrinsic mode excited at the SAW frequency can create modes at other frequencies. We find them in the simulations (Figs. 4-7), but the powers in these modes are orders of magnitude smaller than the power in the dominant modes. We have observed mode hybridization in the past [8]. We do not see radiation due to these modes in the received power spectrum because of two reasons. First, the simulations reveal that these modes usually have lower power than the mode at the SAW frequency (weaker phonon-magnon coupling). Second, they may not be as effective at radiating electromagnetic waves as the mode generated at the SAW frequency (weaker magnon-photon coupling). This antenna is actuated by tripartite coupling between phonons, magnons and photons.

We also observe considerable "anisotropy", i. e. a difference between orientation 1 and orientation 2. This is not surprising given that the array of nanomagnets have in-built anisotropy. Along the x-axis (see Fig. 1), the spacing between the nanomagnet edges is ~25 nm, whereas along the y-axis, it is ~60 nm. The major axes of the ellipse is also the easy magnetization axis and it is along the *x*-axis. The minor axis of the ellipse in the hard magnetization axis and it is along the *y*-axis. Perhaps most importantly, the dipole coupling between the nanomagnets is different in the two directions. Dipole coupling would favor ferromagnetic ordering along the rows (*x*-axis) where nearest neighbors prefer parallel magnetization orientations, whereas along the columns *(y*-axis) anti-ferromagnetic ordering where nearest neighbors prefer antiparallel magnetization orientations. These differences account for the observed anisotropy. Anisotropy effects have also been discussed elsewhere [12].

## III. CONCLUSIONS

In this work, we have demonstrated a novel nano-antenna consisting of magnetostrictive nanomagnets delineated on a piezoelectric substrate. A surface acoustic wave launched in the substrate generates periodic strain the nanomagnets, thereby exciting spin waves via phonon-magnon coupling. These spin waves radiate electromagnetic waves because of magnon-photon coupling. The antenna is thus implemented via tripartite phonon-magnon-photon coupling. Owing to this unusual excitation mechanism, the gain and radiation efficiency of this antenna are not constrained by the usual limits. We have measured these quantities and found them to exceed the theoretical limits of traditional electromagnetic antennas by more than two orders of magnitude. Consequently, they are ideal candidates for ultra-miniaturized antennas (much smaller than the radiated wavelength) needed for embedded applications.

## IV. EXPERIMENTAL SECTION

**Nanomagnet array fabrication**: The $LiNbO_3$ substrate on which the magnetostrictive nanomagnets are fabricated is first cleaned in acetone and isopropyl alcohol and the Al electrodes for launching the SAW are delineated using optical lithography. After delineation of the electrodes, the substrate is spin-coated (spinning rate ~2500 rpm) with a single layer polymethyl methacrylate (PMMA) resist and subsequently baked at 110 °C for 2 min. Next, electron beam lithography is performed using a Raith Voyager Electron Beam Lithography system having accelerating voltage of 50 kV and beam current of 300 pA to open windows for deposition of the nanomagnets. The resists are finally developed in methyl isobutyl ketone and isopropyl alcohol (MIBK-IPA, 1 : 3) for 60 s, which is followed by a cold IPA rinse. A 5 nm-thick Ti adhesion layer is deposited on the patterned substrate using electron beam evaporation base pressure $2.3 \times 10^{-7}$ Torr, followed by the electron beam deposition of 6 nm-thick Co. The lift-off is carried out by remover PG solution.

**Antenna measurements:** Measurements of antenna radiation are carried out with a horn antenna connected to a spectrum analyzer. The setup is surrounded by RF absorbers.

**SUPPORTING INFORMATION**

Supporting information is available from Wiley Online Library or from the corresponding author.

**Acknowledgements**

A. B. and S. B. acknowledge support from the Indo-US Science and Technology Fund Center grant "Center for Nanomagnetics for Energy-Efficient Computing, Communications and Data Storage" (IUSSTF/JC-030/2018).

**Conflict of Interest**

The authors declare no conflict of interest.

**Author contributions**

R. F. fabricated all the samples used in this study. R. F. and J. L. acquired the antenna data and made all the microwave measurements. S. M. carried out the simulations. E. T. supervised the acquisition of the antenna data. A. B. supervised and verified the simulation results. S. B. conceived of the idea and supervised the project. S. B. wrote the manuscript with inputs from all other authors.